\documentclass[twocolumn,superscriptaddress,amsmath, amssymb, amsfonts,preprintnumbers,aps,prd,longbibliography,nofootinbib]{revtex4-1}

\renewcommand{\sec}[1]{\textit{#1. --- }}

\usepackage{graphicx}
\usepackage{dcolumn}
\usepackage{bm}
 \usepackage{amstext}
 \usepackage{amssymb}
\usepackage{subcaption}
 \usepackage{amsmath}
 \usepackage{graphicx}
 \usepackage{color}
 \usepackage{bbold}
 \usepackage{delimset} 
\usepackage[colorlinks=true]{hyperref}
\usepackage{orcidlink}

\usetikzlibrary{decorations.pathmorphing}
\usetikzlibrary{decorations.markings}
\usetikzlibrary{positioning, shapes, snakes, arrows}
 \usepackage{tikz-feynman}

\tikzset{ 
	graviton/.style={line width=.8pt, -latex,decorate, decoration={snake, segment length=4pt,amplitude=1.8pt, pre length=.1cm, post length=.25cm}},
	worldline/.style={gray, line width=1pt},
	worldlineBold/.style={black, line width=.6pt},
        background/.style={black,dotted,line width=1pt},
	zUndirected/.style={line width=1pt},
	zParticle/.style={line width=1pt,postaction={decorate},decoration={markings,mark=at position .6 with {\arrow[#1]{latex}}}},
	zParticleF/.style={line width=1pt,postaction={decorate}},
	cscalar/.style={line width=1pt,postaction={decorate},decoration={markings,mark=at position .6 with {\arrow[#1]{latex}}}},
	cscalar2/.style={line width=1pt,postaction={decorate},decoration={markings,mark=at position .8 with {\arrow[#1]{latex}}}},
	photon/.style={line width =.8pt, decorate, decoration={snake, segment length=3pt, amplitude=1.8pt,  pre length=.1cm, post length=.1cm}},
	 mid arrow/.style={postaction={decorate,decoration={
        markings,
        mark=at position .5 with {\arrow[#1]{latex}}}}} ,
        worlddot/.style={dotted, line width=.8pt},
	worlddot2/.style={dotted, line width=1pt}   }

\DeclareFontFamily{OT1}{pzc}{}
\DeclareFontShape{OT1}{pzc}{m}{it}{<-> s * [1.350] pzcmi7t}{}
\DeclareMathAlphabet{\mathpzc}{OT1}{pzc}{m}{it}

\def\eps{\epsilon}

\def\d{\mathrm{d}}

\def\eE{\mathrm{E}}
\def\eK{\mathrm{K}}

\def\dd{\delta}

\def\d{\mathrm{d}}
\def\eps{\epsilon}

\def\nn{\nonumber}

\def\Eqn#1{Eq.~\eqref{#1}}

\def\Fig#1{Fig.~{\ref{#1}}}

\def\Rcite#1{Ref.~\cite{#1}}
\def\Rcites#1{Refs.~\cite{#1}}

\newcommand{\vev}[1]{\langle #1\rangle}

\newcommand{\be}{\begin{equation}}
\newcommand{\ee}{\end{equation}}
\newcommand{\ba}{\begin{align}}
\newcommand{\ea}{\end{align}}
\ifx\genfrac\sdflkaj\else\fi
\newcommand{\sfrac}[2]{{\textstyle\frac{#1}{#2}}}

\newcommand{\mO}{\mathcal{O}}
\newcommand{\mn}{{\mu\nu}}

\newcommand{\pat}{\partial}

\newcommand*{\vct}[1]{\boldsymbol{#1}}
\newcommand{\gam}{\gamma}
\newcommand{\pin}{p_{\infty}}
\newcommand{\Del}{\Delta}
\newcommand{\Gam}{\Gamma}

\begin{document}

\preprint{HU-EP-23/63-RTG}

\title{Tidal effects and renormalization at fourth post-Minkowskian order}

\author{Gustav Uhre Jakobsen\,\orcidlink{0000-0001-9743-0442}} 
\affiliation{%
Institut f\"ur Physik und IRIS Adlershof, Humboldt-Universit\"at zu Berlin,
Zum Gro{\ss}en Windkanal 2, 12489 Berlin, Germany
}
 \affiliation{Max Planck Institut f\"ur Gravitationsphysik (Albert Einstein Institut), Am M\"uhlenberg 1, 14476 Potsdam, Germany}

\author{Gustav Mogull\,\orcidlink{0000-0003-3070-5717}}
\affiliation{%
Institut f\"ur Physik und IRIS Adlershof, Humboldt-Universit\"at zu Berlin,
Zum Gro{\ss}en Windkanal 2, 12489 Berlin, Germany
}
 \affiliation{Max Planck Institut f\"ur Gravitationsphysik (Albert Einstein Institut), Am M\"uhlenberg 1, 14476 Potsdam, Germany}
 
 \author{Jan Plefka\,\orcidlink{0000-0003-2883-7825}} 
\affiliation{%
Institut f\"ur Physik und IRIS Adlershof, Humboldt-Universit\"at zu Berlin,
Zum Gro{\ss}en Windkanal 2, 12489 Berlin, Germany
}

\author{Benjamin Sauer\,\orcidlink{0000-0002-2071-257X}} 
\affiliation{%
Institut f\"ur Physik und IRIS Adlershof, Humboldt-Universit\"at zu Berlin,
Zum Gro{\ss}en Windkanal 2, 12489 Berlin, Germany
}

\begin{abstract}
We determine the adiabatic tidal contributions to the radiation reacted momentum impulse $\Delta p_i^\mu$ and scattering angle $\theta$ between two scattered massive bodies (neutron stars)
at next-to-next-to-leading post-Minkowskian (PM) order. 
The state-of-the-art three-loop (4PM) 
worldline quantum field theory toolkit using dimensional regularization is employed to establish
the classical observables. We encounter divergent terms in the 
gravito-electric and gravito-magnetic quadrupolar
sectors necessitating the addition of post-adiabatic counterterms in this 
classical theory. This leads us to include also the leading post-adiabatic tidal
contributions to the observables.
The resulting renormalization group flow of the associated post-adiabatic
Love numbers is established
and shown to agree with a recent gravito-electric third post-Newtonian analysis in the non-relativistic limit.
\end{abstract}
 
\maketitle 

With todays routine detection of gravitational waves by the
LIGO-Virgo-Kagra observatories emitted from
binary merger events of black holes and neutron stars in our universe \cite{LIGOScientific:2016aoc,LIGOScientific:2017vwq,LIGOScientific:2021djp} we are 
in the era of gravitational wave astronomy. The upcoming space- and earth-based
third generation of observatories will widen the frequency range and dramatically
increase the sensitivity of the observations \cite{LISA:2017pwj,Punturo:2010zz,Ballmer:2022uxx}. This situation calls for in par precision predictions from theory for the observables
in the gravitational two-body problem. To achieve this a combination of analytical and
numerical approaches is being pursued actively: from the perturbative, analytical side the
post-Newtonian \cite{Blanchet:2013haa,Porto:2016pyg,Levi:2018nxp} and
post-Minkowskian (PM) \cite{Kosower:2022yvp,Bjerrum-Bohr:2022blt,Buonanno:2022pgc,DiVecchia:2023frv,Jakobsen:2023oow}
expansions cover the inspiral phase where the two bodies are still well separated
and weak gravitational fields apply; while the
self-force expansion \cite{Mino:1996nk,Poisson:2011nh,Barack:2018yvs,Gralla:2021qaf}
assumes a mass-hierarchy in the two bodies but works exactly in Newton's coupling $G$.
These perturbative results may be resummed using effective-one-body techniques \cite{Buonanno:1998gg,Buonanno:2000ef}
to extend their validity close to merger
where numerical relativity (NR) \cite{Pretorius:2005gq,Boyle:2019kee,Damour:2014afa} techniques 
become indispensable. 
 
Recently, considerable progress has been made  upon importing modern techniques
from perturbative quantum field theory (QFT) to the problem in the PM expansion. 
While the natural habitat for the PM expansion is the scattering
of black holes or neutron stars
\cite{Kovacs:1978eu,Westpfahl:1979gu,Bel:1981be,Damour:2017zjx,Hopper:2022rwo}, 
the scattering data may nevertheless be used to inform models for the bound-state problem
that should become particularly relevant for highly eccentric orbits
 \cite{Cheung:2018wkq,Kalin:2019rwq,Kalin:2019inp,Saketh:2021sri,Gonzo:2023goe,Cho:2021arx}. 
As long as the objects' separation is large compared to their intrinsic sizes, 
they have an effective description in terms of a massive point particle
coupled to Einstein's theory of gravity that may be systematically corrected for
intrinsic degrees of freedom such as spin or tidal effects 
\cite{Goldberger:2004jt}.
Based on this effective worldline approach two-body scattering observables (deflections and
Bremsstrahlung waveforms)
have recently been computed up to next-to-next-to-next-to leading order (deflections)
and leading order (Bremsstrahlung) 
\cite{Kalin:2020mvi,Kalin:2020fhe,Kalin:2020lmz,Dlapa:2021npj,Dlapa:2021vgp,
Liu:2021zxr,Mougiakakos:2021ckm,Riva:2021vnj,Mougiakakos:2022sic,Riva:2022fru,
Mogull:2020sak,Jakobsen:2021smu,Jakobsen:2021lvp,Jakobsen:2021zvh,Jakobsen:2022fcj,
Jakobsen:2022psy,Shi:2021qsb,Bastianelli:2021nbs,Comberiati:2022cpm,Wang:2022ntx}.
In parallel, great leaps in the QFT based PM expansions were achieved using techniques based on scattering amplitudes in which quantum field act as avatars of BHs or NSs
\cite{Neill:2013wsa,Luna:2017dtq,Kosower:2018adc,Cristofoli:2021vyo,Bjerrum-Bohr:2013bxa,Bjerrum-Bohr:2018xdl,Bern:2019nnu,Bern:2019crd,Bjerrum-Bohr:2021wwt,Cheung:2020gyp,Bjerrum-Bohr:2021din,DiVecchia:2020ymx,DiVecchia:2021bdo,DiVecchia:2021ndb,DiVecchia:2022piu,Heissenberg:2022tsn,Damour:2020tta,Herrmann:2021tct,Damgaard:2019lfh,Damgaard:2019lfh,Damgaard:2021ipf,Damgaard:2023vnx,Aoude:2020onz,AccettulliHuber:2020dal,Brandhuber:2021eyq,Bern:2022kto,Aoude:2022thd,Aoude:2023vdk,Bautista:2023szu,Damgaard:2023ttc,Brandhuber:2023hhy,Herderschee:2023fxh,Georgoudis:2023lgf,Elkhidir:2023dco,Caron-Huot:2023vxl}.

Next to the masses and spins of the compact objects, tidal deformations are a significant
astrophysical phenomenon and observational goal. Neutron stars (NSs) develop a quadrupole moment due to the
tidal interaction with their companion star or black hole (BH) \cite{Flanagan:2007ix,Hinderer:2007mb}. 
The strength of this effect is parametrized by the Love numbers that 
quantify the magnitude of the induced multipole moment in response to an
external gravitational field. Measuring them through gravitational waves
provides insights into the strong interaction matter within neutron stars.
In fact, the gravitational wave signal GW170817 observing the first 
NS-NS merger \cite{LIGOScientific:2017vwq}
was able to put constraints on the first (gravito-electric) Love
number with consequences for the neutron star equation of 
state \cite{LIGOScientific:2018hze,LIGOScientific:2018cki,Chatziioannou:2020pqz,Pradhan:2022rxs}.
The tidal interactions give rise to oscillation modes of the NS, and in particular the
so-called $f$-mode dynamical tides \cite{Will:1983vlw,Steinhoff:2016rfi} have been argued to be central for inferring the NS equation of state from the emitted gravitational waves in a merger \cite{Pratten:2021pro}.
In the adiabatic limit the tides do not oscillate independently 
and are locked to the external gravito-electric and gravito-magnetic fields --- 
the limit we shall consider in this work. 

In the PN expansion gravito-magnetic and gravito-electric tides have been established up to
2PN order \cite{Bini:2012gu,Henry:2019xhg} and recently
the state-of-the art has been extended to 3PN for dynamical
and adiabatic tides \cite{Saketh:2023bul,Mandal:2023hqa}. Working in dimensional regularization
the authors of \Rcite{Mandal:2023hqa} encountered a UV divergence that necessitated the
inclusion of a post-adiabatic counterterm, leading to a renormalization group flow
of its post-adiabatic Love number.
Meanwhile, in the PM expansion
the two-body scattering observables of the impulse (change of momentum) and
scattering angle in the presence of tidal interactions 
have been determined at 2PM \cite{Bini:2020flp,Cheung:2020sdj,Haddad:2020que,Bern:2020uwk}
and 3PM \cite{Kalin:2020lmz} order in the conservative sector.
We updated the 3PM result to include dissipation \cite{Jakobsen:2022psy} using 
the in-in worldline quantum field theory (WQFT) formalism. In addition, the Bremsstrahlung waveform
with tidal effects at leading order was established in \Rcites{Mougiakakos:2022sic,Jakobsen:2022psy}.

In this Letter we lift
this tidal precision prediction for the impulse and scattering angle to the 
4PM, i.e.~next-to-next-to-leading, order in the
PM expansion --- both in the conservative and dissipative sectors
and for the gravito-electric and gravito-magnetic tides. As it turns out, this classical computation suffers
from an UV (ultra-violet) divergence that may be attributed to the point-particle 
approximation of the neutron star, equivalent to what was seen at 3PN order \cite{Saketh:2023bul,Mandal:2023hqa}\footnote{This divergence was also seen in gravitational radiation from a single compact object with a 
quadrupole in Refs.~\cite{Blanchet:1997jj,Goldberger:2009qd}.}.
We regulate the theory using dimensional regularization \emph{in the bulk}, i.e.~the
worldline action remains one dimensional. While Newton's constant $G$ is continued to
$D=4-2\epsilon$ dimensions,  $G_{D}=(  4\pi e^{\gamma_{E}}R^{2})^{-\epsilon} G$, introducing an
arbitrary length scale $R$, the Love numbers are not dimensionally continued. Removing
the UV divergence through a post-adiabatic counterterm then induces a renormalization
group flow of the associated post-adiabatic Love numbers, matching the flow in the
gravito-electric sector in the PN analysis of Ref.~\cite{Mandal:2023hqa}. 
Using our results for the impulse and scattering angle we also establish the
linear and angular momentum, at 4PM and 3PM order respectively,
radiated off by the gravitational waves emitted in the encounter of the two neutron stars (NSs).
 
\sec{Worldline effective action}
The effective description of non-spinning compact objects (neutron stars) including the leading-order
adiabatic tidal couplings takes the form of a point-particle action
$
S=\sum_{i=1}^{2}S^{(i)}_{\rm pp} +S^{(i)}_{\rm tidal}
$, where \cite{Kalin:2020mvi}
\begin{align}\label{eq:pp}
	S^{(i)}_{\rm pp} &\!=\!-m_i \int\!\d\tau \left[\frac{1}{2e} g_{\mu\nu}\dot x_i^{\mu}\dot x_i^{\nu} + \frac{e}{2}
	\right ]\,,  \\ \label{eq:tidal}
	S^{(i)}_{\rm tidal} &= - m_i \!\int\!\d\tau \!\left[
	\frac{c_{E^2}^{(i)}}{e^{3}} E_{\mu \nu}^{(i)} E^{(i) \mu \nu}
	\!+\!\frac{c_{B^2}^{(i)}}{e^{3}} B_{\mu \nu}^{(i)} B^{(i) \mu \nu}\right]\!\!. 
\end{align}
Here $x^{\mu}_{i}(\tau)$ is the trajectory of the $i$th body of mass $m_{i}$ and $e(\tau)$ is the
einbein ensuring reparametrization invariance of the worldline theory. 
The quadrupole Love numbers  $c^{(i)}_{E^2}$ and $c^{(i)}_{B^2}$ (Wilson
coefficients in an effective field theory nomenclature) are of mass dimension
-4 and  couple to the gravito-electric and gravito-magnetic curvature tensors
\begin{align}
  E_{\mu \nu}^{(i)}&:= R_{\mu \alpha \nu \beta} \dot{x}_i^{ \alpha} \dot{x}_i^{ \beta}\,,&
  B_{\mu \nu}^{(i)}&:= R^{*}_{\mu \alpha \nu \beta} \dot{x}_i^{ \alpha} \dot{x}_i^{ \beta}\,,
\end{align}
with the dual Riemann tensor
$R^*_{\mu \alpha \nu \beta}:= \sfrac{1}{2} \epsilon_{\nu \beta \rho \sigma} {R_{\mu \alpha}}^{\rho \sigma}$. Introducing
$ B_{\mu \nu \rho}^{(i)}= R_{\alpha \mu \nu \rho} \dot{x}_i^{ \alpha} \sqrt{\dot{x}_i^2}$
we note the relation
\be\label{eq:generalizeToD}
B_{\mu \nu}^{(i)} B^{(i) \mu \nu} = 
 E_{\mu \nu}^{(i)} E^{(i) \mu \nu} -
 \frac{1}{2}B_{\mu \nu \rho}^{(i)}B^{(i)\,\mu \nu \rho} \,,
\ee
that generalizes \eqref{eq:tidal} to $D$ dimensions.
These are the first of an infinite series of tidal Love number couplings,
capturing the linear response of the compact body to an external gravitational field.
For the case of a black hole they are known to vanish~\cite{Fang:2005qq,Damour:2009vw,Binnington:2009bb}.

The two neutron stars $x_{i}^{\mu}(\tau)$ interact gravitationally according to the gauge-fixed Einstein-Hilbert action
\begin{align}\label{eq:einsteinHilbert}
S_{\rm bulk}&=\int\!\d^Dx\left (\! -\frac{1}{16\pi G_{D}}\sqrt{-g}R\,
+(\partial_\nu h^{\mu\nu}-\sfrac12\partial^\mu{h^\nu}_\nu)^2 \!\right )
\end{align}
in the bulk, where $g_{\mu\nu}=\eta_{\mu\nu} + \sqrt{32\pi G_{D}} h_{\mu\nu}$
and $h_{\mu\nu}$ is the graviton field. 
Here,  we take 
$G_{D}=(  4\pi e^{\gamma_{E}}R^{2})^{-\epsilon} G$, the extension of Newton's constant
$G$ to $D=4-2\epsilon$ dimensions, working in an $\overline{\text{MS}}$ scheme adapted to
configuration space.
$R$ denotes an intrinsic length scale
of the compact object, such as its radius.

Let us briefly comment on our PM counting scheme.
For a neutron star, or other compact body whose radius is of the order of its Schwarzschild radius, it is natural to factor the scale of the Schwarzschild radius out of the Love numbers such that
\begin{align}
  c_{E^2}^{(i)}&=(Gm_i)^4 \tilde c_{E^2}^{(i)}\,,&
  c_{B^2}^{(i)}&=(Gm_i)^4 \tilde c_{B^2}^{(i)}\,,
\end{align}
with dimensionless Love numbers $\tilde c_{E^2/B^2}^{(i)}$ of order unity.
From that perspective, the results reported in this Letter are enhanced with an additional factor of $G^4$,
which pushes them to the (physical) 8PM order.
In this Letter, however, we opt for a (formal) PM counting aligned with all explicit instances of $G$ in the action ---
 Eqs.~\eqref{eq:einsteinHilbert} and~\eqref{eq:ct} below.
 For the adiabatic tidal results this implies that $n$PM corresponds with $(n-1)$ loops while for the post-adiabatic interaction introduced below $n$PM corresponds with $(n-3)$ loops
 (see e.g. also the discussion in Ref.~\cite{Rettegno:2023ghr} regarding PM counting with spin).
 
 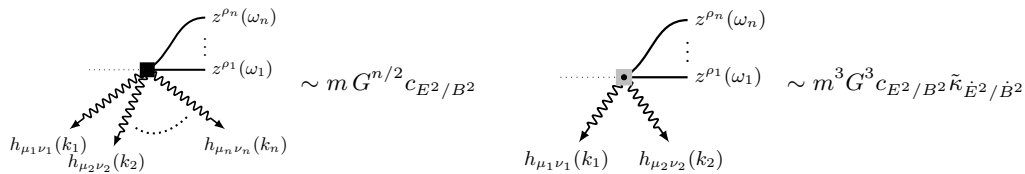
\begin{figure*}[th!]
$\displaystyle
\resizebox{0.22\textwidth}{!}{
		  \begin{tikzpicture}[baseline={(current bounding box.center)}]
		\coordinate (in) at (-1,0);
		\coordinate (out1) at (1,0);
		\coordinate (out2) at (1,0.5);
		\coordinate (out3) at (1,0.9);
		\coordinate (x) at (0,0);
		\coordinate (y2) at (-0.2,-1.0);
		\coordinate (y4) at (0.7,-0.8);
		\node (k1) at (-1.7,-1.3) {$h_{\mu_1\nu_1}(k_1)$};
		\node (k2) at (-0.7,-1.6) {$h_{\mu_2\nu_2}(k_2)$};
		\node (k4) at (1.7,-1.3) {$h_{\mu_n\nu_n}(k_n)$};
		\draw (out1) node [right] {$z^{\rho_1}(\omega_1)$};
		\draw (out2) node [right] {$\!\!\!\vdots$};
		\draw (out3) node [right] {$z^{\rho_n}(\omega_n)$};
		\draw [dotted] (in) -- (x);
		\draw [zUndirected] (x) -- (out1);
		\draw [zUndirected] (x) to[out=30,in=180] (out3);
		\draw [graviton] (x) -- (k1);
		\draw [graviton] (x) -- (k2);
		\draw [dotted, line width=1pt] (y2)  to[out=-35,in=-135] (y4);
		\draw [graviton] (x) -- (k4);
		\draw [fill] (x) +(-.12,-.12) rectangle ++(.12,.12);
		\end{tikzpicture}}
		\sim m \,G^{n/2} c_{E^{2}/B^{2}}
$
		\quad
		$\displaystyle
		\resizebox{0.2\textwidth}{!}{
		\begin{tikzpicture}[baseline={(current bounding box.center)}]
	  \coordinate (in) at (-1,0);
	  \coordinate (out1) at (1,0);
	  \coordinate (out2) at (1,0.5);
	  \coordinate (out3) at (1,0.9);
	  \coordinate (x) at (0,0);
	  \coordinate (xd) at (0,0.25);
	  \node (k1) at (-.9,-1.3) {$h_{\mu_1\nu_1}(k_1)$};
	  \node (k2) at (.9,-1.3) {$h_{\mu_2\nu_2}(k_2)$};
	  \draw (out1) node [right] {$z^{\rho_1}(\omega_1)$};
	  \draw (out2) node [right] {$\!\!\!\vdots$};
	  \draw (out3) node [right] {$z^{\rho_n}(\omega_n)$};
	  \draw [dotted] (in) -- (x);
	  \draw [zUndirected] (x) -- (out1);
	  \draw [zUndirected] (x) to[out=30,in=180] (out3);
	  \draw [graviton] (x) -- (k1);
	  \draw [graviton] (x) -- (k2);
	  \draw [fill, color=lightgray] (x) +(-.12,-.12) rectangle ++(.12,.12);
	  \draw [fill] (x) circle (0.04);
	  \end{tikzpicture} }\sim m^{3} G^{3} c_{E^{2}/B^{2}} \tilde{\kappa}_{{\dot E}^{2}/{\dot B}^{2}}
	  $
	  \caption{The tidal interaction vertices needed for the 4PM (order $\kappa^{8}$) 
	  computation, originating from $S_{\text{tidal}}$ \eqref{eq:tidal} and the counter-term vertices originating from $S_{\text{ct}}$ \eqref{eq:ct}.
	  }
	  \label{fig:tidalvertices}
	  \end{figure*}	  
\sec{Renormalization}
As mentioned above, in the computation of the impulse at the (formal) 4PM level
one encounters an UV divergence that
 is canceled upon including the
post-adiabatic tidal counter term
\begin{align}\label{eq:ct}
\begin{aligned}
S^{(i)}_{\rm ct} = - m_i^{3} G^{2}\int\!\d\tau \Bigl[&
	\frac{c_{E^{2}}^{(i)}{\tilde \kappa}_{\dot{E}^2}^{(i)}}{e^{5}} {\dot E}_{\mu \nu}^{(i)} {\dot E}^{(i) \mu \nu} \\
	&+\frac{c_{B^{2}}^{(i)}{\tilde \kappa}_{\dot{B}^2}^{(i)}}{e^{5}} {\dot B}_{\mu \nu}^{(i)} {\dot B}^{(i) \mu \nu}\Bigr]
\end{aligned}
\end{align}
to the total action.
Note that we need to use the 4D Newton constant $G$ here, as
the worldline action remains one dimensional.
In the present work, the post-adiabatic counter term \Eqn{eq:ct} is required only at tree-level.
At leading order in $G$ its last term straightforwardly generalizes
to $D$-dimensions as
\be
\dot B_{\mu \nu}^{(i)} \dot B^{(i) \mu \nu} = 
 \dot E_{\mu \nu}^{(i)} \dot E^{(i) \mu \nu} -
 \frac{1}{2}\dot B_{\mu \nu \rho}^{(i)}\dot B^{(i)\,\mu \nu \rho}
 +
 \mO(G^3)\,.
\ee
In \Eqn{eq:ct} the dimensionless post-adiabatic Love numbers $\tilde \kappa_{\dot{E}^2/\dot{B}^2}$  take the form (dropping the neutron
star label $(i)$)
\begin{align}\begin{split}
\tilde{\kappa}_{\dot{E}^2} &= -\frac{107}{105}  \frac{1}{\epsilon} + \kappa_{\dot{E}^2}\, ,
\\
\tilde{\kappa}_{\dot{B}^2} &= -\frac{107}{105} \frac{1 }{\epsilon} + \kappa_{\dot{B}^2}\, ,
\end{split}
\end{align}
with the counter-terms removing the divergences appearing at the 4PM order
being given by the $1/\epsilon$ terms. We also include finite post-adiabatic Love numbers
$\kappa_{\dot{E}^2}$ and  $\kappa_{\dot{B}^2}$. They experience a renormalization group
flow as follows: the bare gravitational coupling $G_{D}$ is  independent of the scale $R$, so
the flow equation for Newton's constant reads
\begin{align}
\begin{aligned}
\label{RGEG}
0&= R\frac{\d}{\d R}G_{D}=  R\frac{\d}{\d R} \Bigl [( 4\pi  e^{\gamma_{E}}R^{2})^{-\epsilon} G  \Bigr ]\\
& = \Bigl [ -2\epsilon G + R\frac{\d}{\d R}G\Bigr ] ( 4\pi  e^{\gamma_{E}}R^{2})^{-\epsilon}\,,
\end{aligned}
\end{align}
i.e.~there is no flow of $G$ in $D=4$ dimensions. 
The bare couplings in \eqref{eq:tidal} and \eqref{eq:ct} do not depend on the scale $R$, hence
\begin{align}
\label{RGEcomp}
0 &=R\frac{\d}{\d R}c_{E^{2}} \nn \\
0 &=R\frac{\d}{\d R}( c_{E^{2}}G^{2}\tilde{\kappa}_{\dot{E}^2})\\ &=
c_{E^{2}}G^{2} R\frac{\d}{\d R}\kappa_{\dot{E}^2} - \frac{107}{105}\nn
 \frac{ c_{E^{2}}}{\epsilon} \, R\frac{\d}{\d R}G^{2}\, .
\end{align}
Together with \eqref{RGEG}, this then
yields the $\beta$-functions for the renormalized couplings $\kappa_{\dot{E}^2}$ and
$\kappa_{\dot{B}^2}$ \cite{Saketh:2023bul,Mandal:2023hqa}
\begin{align}
\beta_{\kappa_{\dot{E}^2}} &=R\frac{\d \kappa_{\dot{E}^2}}{\d R}= \frac{428}{105}\, , &
\beta_{\kappa_{\dot{B}^2}} = R\frac{\d \kappa_{\dot{B}^2}}{\d R}=\frac{428}{105}\, .
\end{align}
These induce a logarithmic flow of the renormalized post-adiabatic Love numbers as
\begin{subequations}
\begin{align}
\kappa_{\dot E^{2}}(R) &= \kappa_{\dot E^{2}}(R_{0}) + \frac{428}{105}\,\log\left[
\frac{R}{R_{0}}\right ]\,,\\
\kappa_{\dot B^{2}}(R) &= \kappa_{\dot B^{2}}(R_{0}) + \frac{428}{105}\,\log\left[
\frac{R}{R_{0}}\right ]\,,
\end{align}
\end{subequations}
where $R_0$ is an arbitrary length scale.\footnote{Note that the 3PN reference
\cite{Mandal:2023hqa} works with different conventions than we do. Their post-adiabatic
coupling is minus one-half ours, $\kappa_{PN}= - \kappa/2$, their adiabatic coupling 
$\lambda_{PN}=4m c_{E^{2}}$ and they take $D-1=3+\varepsilon_{PN}$, i.e. $\varepsilon_{PN}=-2\epsilon$.
Taking this into account, we agree with their findings.}


\begin{figure*}[t]
  \centering
    \centering
    \begin{tikzpicture}[baseline={(current bounding box.center)},scale=.7]
  		\coordinate (inA) at (0.4,.6);
  		\coordinate (outA) at (3.6,.6);
  		\coordinate (inB) at (0.4,-.6);
  		\coordinate (outB) at (3.6,-.6);
  		\coordinate (xA) at (1,.6);
  		\coordinate (xyA) at (1.5,.6);
  		\coordinate (yA) at (2,.6);
  		\coordinate (yzA) at (1.5,.6);
  		\coordinate (zA) at (3,.6);
  		\coordinate (xB) at (0.8,-.6);
		\coordinate (xyB) at (1.5,-.6);
  		\coordinate (yB) at (2.2,-.6);
  		\coordinate (zB) at (3,-.6);
  		\draw [fill] (xyA) +(-.12,-.12) rectangle ++(.12,.12);
  		\draw [fill] (zA) circle (.08);
  		\draw [fill] (xB) circle (.08);
  		\draw [fill] (yB) circle (.08);
  		\draw [fill] (zB) circle (.08);
		\draw [fill] (xyB) circle (.08);
  		\draw [dotted] (inA) -- (outA);
  		\draw [dotted] (inB) -- (outB);
  		\draw [zParticleF] (zA) -- (outA);
  		\draw [draw=none] (xA) to[out=40,in=140] (zA);
  		\draw [zParticleF] (xyA) -- (yA);
  		\draw [zParticleF] (yA) -- (zA);
  		\draw [photon] (xyA) -- (xB);
		\draw [photon] (xyA) -- (xyB);
  		\draw [photon] (xyA) -- (yB);
  		\draw [photon] (zA) -- (zB);
  	\end{tikzpicture}
	\quad
     \begin{tikzpicture}[baseline={(current bounding box.center)},scale=.7]
  		\coordinate (inA) at (0.4,.6);
  		\coordinate (outA) at (3.6,.6);
  		\coordinate (inB) at (0.4,-.6);
  		\coordinate (outB) at (3.6,-.6);
  		\coordinate (xA) at (0.8,.6);
		\coordinate (xxA) at (1.5,.6);
  		\coordinate (xyA) at (1.5,.6);
  		\coordinate (yA) at (2,.6);
  		\coordinate (yzA) at (2.5,.6);
  		\coordinate (zA) at (3,.6);
  		\coordinate (xB) at (0.8,-.6);
		\coordinate (xxB) at (1.5,-.6);
  		\coordinate (yB) at (2,-.6);
  		\coordinate (zB) at (3,-.6);
  		\draw [fill] (xA) circle (.08);
  		\draw [fill] (yzA) +(-.12,-.12) rectangle ++(.12,.12);
  		\draw [fill] (xB) circle (.08);
  		\draw [fill] (yB) circle (.08);
  		\draw [fill] (zB) circle (.08);
		\draw [fill] (xxA) circle (.08);
		\draw [fill] (xxB) circle (.08);
  		\draw [dotted] (inA) -- (outA);
  		\draw [dotted] (inB) -- (outB);
  		\draw [zParticleF] (zA) -- (outA);
  		\draw [draw=none] (xA) to[out=40,in=140] (zA);
  		\draw [zParticleF] (xA) -- (yA);
  		\draw [zParticleF] (yA) -- (zA);
  		\draw [photon] (xA) -- (xB);
		\draw [photon] (xxA) -- (xxB);
  		\draw [photon] (yzA) -- (yB);
  		\draw [photon] (yzA) -- (zB);
  	\end{tikzpicture}
	\quad
    \begin{tikzpicture}[baseline={(current bounding box.center)},scale=.7]
  		\coordinate (inA) at (0.4,.6);
  		\coordinate (outA) at (3.6,.6);
  		\coordinate (inB) at (0.4,-.6);
  		\coordinate (outB) at (3.6,-.6);
  		\coordinate (xA) at (1,.6);
  		\coordinate (xyA) at (1.5,.6);
  		\coordinate (yA) at (2,.6);
  		\coordinate (yzA) at (1.5,.6);
  		\coordinate (zA) at (3,.6);
  		\coordinate (xB) at (1,-.6);
  		\coordinate (yB) at (1.7,-.6);
		\coordinate (yyB) at (2.3,-.6);
  		\coordinate (zB) at (3,-.6);
  		\draw [fill] (yA)+(-.12,-.12) rectangle ++(.12,.12);
  		\draw [fill] (xB) circle (.08);
  		\draw [fill] (yB) circle (.08);
		\draw [fill] (yyB) circle (.08);
  		\draw [fill] (zB) circle (.08);
  		\draw [dotted] (inA) -- (outA);
  		\draw [dotted] (inB) -- (outB);
  		\draw [zParticleF] (zA) -- (outA);
  		\draw [draw=none] (xA) to[out=40,in=140] (zA);
  		\draw [zParticleF] (yA) -- (zA);
  		\draw [photon] (yA) -- (xB);
  		\draw [photon] (yA) -- (yB);
		\draw [photon] (yA) -- (yyB);
  		\draw [photon] (yA) -- (zB);
  	\end{tikzpicture}
	\quad
    \begin{tikzpicture}[baseline={(current bounding box.center)},scale=.7]
  		\coordinate (inA) at (0.4,.6);
  		\coordinate (outA) at (3.6,.6);
  		\coordinate (inB) at (0.4,-.6);
  		\coordinate (outB) at (3.6,-.6);
  		\coordinate (xA) at (1,.6);
  		\coordinate (xyA) at (1.5,.6);
  		\coordinate (xy0) at (1.5,0.2);
  		\coordinate (z0) at (3,0);
  		\coordinate (yA) at (2,.6);
  		\coordinate (yzA) at (1.5,.6);
  		\coordinate (zA) at (3,.6);
  		\coordinate (xB) at (0.8,-.6);
		\coordinate (xxB) at (1.5,-.6);
  		\coordinate (yB) at (2.2,-.6);
  		\coordinate (zB) at (3,-.6);
  		\draw [fill] (xy0) circle (.08);
  		\draw [fill] (zA) +(-.12,-.12) rectangle ++(.12,.12);
  		\draw [fill] (xB) circle (.08);
		\draw [fill] (xxB) circle (.08);
  		\draw [fill] (yB) circle (.08);
  		\draw [fill] (zB) circle (.08);
  		\draw [dotted] (inA) -- (outA);
  		\draw [dotted] (inB) -- (outB);
  		\draw [zParticleF] (zA) -- (outA);
  		\draw [photon] (xy0)to[out=0,in=-140] (zA);
  		\draw [draw=none] (xA) to[out=40,in=140] (zA);
  		\draw [photon] (xy0) -- (xB);
		\draw [photon] (xy0) -- (xxB);
  		\draw [photon] (xy0) -- (yB);
  		\draw [photon] (zA) -- (zB);
  	\end{tikzpicture}
    \caption{\small Examples of 4PM graphs linear in tidal coefficients contributing to
    the test-body $m_{1}m_{2}^{4}$ sector.}
  \label{fig:tidal-graphs1}
\end{figure*}
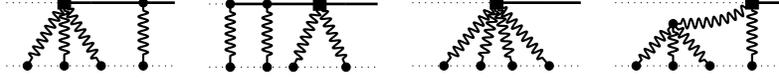


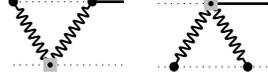
\begin{figure*}[t]
\label{fig:tidal-graphs2}
  \centering
\begin{tikzpicture}[baseline={(current bounding box.center)},scale=.7]
  		\coordinate (inA) at (1.5,.6);
  		\coordinate (outA) at (3.6,.6);
  		\coordinate (inB) at (1.5,-.6);
  		\coordinate (outB) at (3.6,-.6);
  		\coordinate (xA) at (0.8,.6);
		\coordinate (xxA) at (1.5,.6);
  		\coordinate (xyA) at (1.5,.6);
  		\coordinate (yA) at (2,.6);
  		\coordinate (yzA) at (2.2,-.6);
  		\coordinate (zA) at (3,.6);
  		\coordinate (xB) at (0.8,-.6);
		\coordinate (xxB) at (1.5,-.6);
  		\coordinate (yB) at (1.5,.6);
  		\coordinate (zB) at (3,.6);
  		\draw [fill] (yB) circle (.08);
  		\draw [fill] (zB) circle (.08);
  		\draw [dotted] (inA) -- (outA);
  		\draw [dotted] (inB) -- (outB);
  		\draw [zParticleF] (zB) -- (outA);
  		\draw [photon] (yzA) -- (yB);
  		\draw [photon] (yzA) -- (zB);
		\draw [fill, color=lightgray] (yzA) +(-.12,-.12) rectangle ++(.12,.12);
		 \draw [fill]  (yzA) circle (0.04);
  	\end{tikzpicture}
\quad
 \begin{tikzpicture}[baseline={(current bounding box.center)},scale=.7]
  		\coordinate (inA) at (1.5,.6);
  		\coordinate (outA) at (3.6,.6);
  		\coordinate (inB) at (1.5,-.6);
  		\coordinate (outB) at (3.6,-.6);
  		\coordinate (xA) at (0.8,.6);
		\coordinate (xxA) at (1.5,.6);
  		\coordinate (xyA) at (1.5,.6);
  		\coordinate (yA) at (2,.6);
  		\coordinate (yzA) at (2.5,.6);
  		\coordinate (zA) at (3,.6);
  		\coordinate (xB) at (0.8,-.6);
		\coordinate (xxB) at (1.5,-.6);
  		\coordinate (yB) at (1.8,-.6);
  		\coordinate (zB) at (3.2,-.6);
  		\draw [fill] (yB) circle (.08);
  		\draw [fill] (zB) circle (.08);
  		\draw [dotted] (inA) -- (outA);
  		\draw [dotted] (inB) -- (outB);
  		\draw [zParticleF] (yzA) -- (outA);
  		\draw [photon] (yzA) -- (yB);
  		\draw [photon] (yzA) -- (zB);
		\draw [fill, color=lightgray] (yzA) +(-.12,-.12) rectangle ++(.12,.12);
		 \draw [fill]  (yzA) circle (0.04);
  	\end{tikzpicture}
	 \caption{\small The post-adiabatic graphs proportional to
	 $\tilde\kappa_{\dot E^{2}}$ and  $\tilde\kappa_{\dot B^{2}}$ that also cancel the divergence.}
  \label{fig:tidal-graphs2}
\end{figure*}

\sec{Computation} Our 4PM computation is performed using the WQFT three-loop workflow 
as described in
\cite{Jakobsen:2023ndj,Jakobsen:2023hig}, which we briefly review.
The full tidal effective field theory is given by the sum of the bulk \Eqn{eq:einsteinHilbert} and worldline actions
of Eqs.~\eqref{eq:pp} and \eqref{eq:ct}. For the worldline trajectories
we perform a background field expansion about straight line trajectories
\be
x^{\mu}_{i}(\tau) = b_{i}^{\mu}+ v_{i}^\mu \tau + z_{i}^{\mu}(\tau)\, ,
\ee
with perturbative deflections $z_i^\mu(\tau)$. This setup
reflects the scattering scenario parametrized by the impact parameters $b_{i}^{\mu}$ and incoming velocities
$v_{i}^{\mu}$.
In addition we define the physical impact parameter $b^\mu=|b| \hat b^{\mu}=(b_2-b_1)^\mu$ and we impose $v_{i}\cdot b=0$. 
Moreover, in the PM expansion the metric is taken to be
 $g_\mn=\eta_\mn +\sqrt{32\pi G_{D}}h_\mn$.
The goal is to construct perturbative-in-$G$ solutions to the equations of motion for the 
deflections $z_i^\mu(\tau)$, which is efficiently generated in a diagrammatic fashion
in the WQFT formalism --- see \Rcite{Jakobsen:2023oow} for a recent review. 

The propagating fields $z_i^\mu(\tau)$ and $h_{\mu\nu}(x)$ have the retarded propagators
\begin{subequations}
  \begin{align}
    \begin{tikzpicture}[baseline={(current bounding box.center)}]
      \coordinate (in) at (-0.6,0);
      \coordinate (out) at (1.4,0);
      \coordinate (x) at (-.2,0);
      \coordinate (y) at (1.0,0);
      \draw [zUndirected] (x) -- (y) node [midway, below] {$\omega$} node [midway, above] {$\rightarrow$};
      \draw [background] (in) -- (x);
      \draw [background] (y) -- (out);
      \draw [fill] (x) circle (.08) node [above] {$\mu$};
      \draw [fill] (y) circle (.08) node [above] {$\nu$};
    \end{tikzpicture}&=\frac{-i\eta^{\mu\nu}}{m_i(\omega+i0^+)^{2}}\,,
        \\
      \label{eq:gravProp}
      \begin{tikzpicture}[baseline={(current bounding box.center)}]
    \begin{feynman}
    \coordinate (x) at (-.7,0);
    \coordinate (y) at (0.5,0);
    \draw [photon] (x) -- (y) node [midway, below] {$k$} node [midway, above] {$\rightarrow$};
    \draw [fill] (x) circle (.08) node [above] {$\mu\nu$};
    \draw [fill] (y) circle (.08) node [above] {$\rho\sigma$};
    \end{feynman}
    \end{tikzpicture}
          &=\frac{i(\eta_{\mu(\rho}\eta_{\sigma)\nu}-\sfrac1{D-2}\eta_{\mu\nu}\eta_{\rho\sigma})
          }{k^{2}+ \text{sgn}{(k^{0})}i 0^{+}}\,,
  \end{align}
  \end{subequations}
using the Schwinger-Keldysh in-in formalism~\cite{Jakobsen:2022psy,Kalin:2022hph}.
The worldline vertex rules originating from $S_{\rm pp}$ in \Eqn{eq:pp} at lower multiplicities
have been exposed explicitly in \cite{Jakobsen:2021zvh,Jakobsen:2023ndj}: the vertex
couples one graviton to $m$ worldline deflections and conserves the energy on the worldline.
The vertices originating from the tidal terms \Eqn{eq:tidal} involve $n\geq 2$
gravitons and $m$ worldline deflections --- for the 4PM order computation we need the vertices with up to four graviton legs $n=2,3,4$.
The counter-term \eqref{eq:ct} on the other hand, gives rise to an $n\geq 2$-graviton
and $m$-worldline vertex, that we only need for $n=2$ and $m=0,1$ cp.~\Fig{fig:tidalvertices}.
 The bulk graviton vertices are standard -- yet involved.
 Using these Feynman rules the WQFT  tree-level one-point functions
$\vev{z_{i}^{\mu}(\tau)}$ may be systematically constructed.
They solve the classical equations of motion 
\cite{Boulware:1968zz,Jakobsen:2023oow}.

\sec{WQFT workflow at 4PM} For the computation of the impulse, i.e.~the change
of momentum under the scattering process,
\be
\Delta p^{\mu}_{i} = -m_{i} \omega^{2}\vev{z^{\mu}_{i}(\omega)}\Bigl |_{\omega=0}
=\sum_{n>0} \Delta p^{(n)\, \mu}_{i}\,.
\ee
At the 4PM level $\Delta p^{(4)\,\mu}_{i}$ we employ an in-house {\tt FORM}- \cite{Ruijl:2017dtg} 
and {\tt Mathematica}-based code that employs 
a Berends-Giele type recursion for the integrand construction.
At this order we face three-loop Feynman integrals that depend on the momentum transfer $q^{\mu}$ and
the relativistic $\gamma=v_{1}\cdot v_{2}$ factor. The $|q|$ may be scaled out and
we subsequently reduce the single-scale tensor integrals to scalars. 
The integration-by-part (IBP) reduction and projection on the 4PM master integral basis was obtained in \cite{Jakobsen:2023ndj,Jakobsen:2023hig}. 
The 4PM master integrals in turn are three-loop single scale integrals
(depending on $\gamma$) that have
been solved  employing the differential canonical equation method \cite{Gehrmann:1999as,Henn:2013pwa}
and the method of regions \cite{Beneke:1997zp} in the conservative and dissipative domains 
\cite{Jakobsen:2023ndj,Jakobsen:2023hig}, see also 
\cite{Bern:2021yeh,Dlapa:2022lmu,Dlapa:2023hsl,Damgaard:2023ttc}.

The final step is a Fourier transform of the momentum transfer $q^{\mu}$ to
impact parameter space. Here one novelty to the spinning 4PM computation of \cite{Jakobsen:2023ndj,Jakobsen:2023hig} is the appearance of $\log\!|Rq|$ terms.
\begin{align}\label{FTlog}
&\int_q e^{iq\cdot b}\dd(q\cdot v_1)\dd(q\cdot v_2)|q|^\nu\log\!|Rq| \nn\\
&\quad=\frac{2^{\nu-1}}{\pi^{(D-2)/2}\sqrt{\gamma^2-1}}\frac{\Gamma(\frac{D-2+\nu}2)}{\Gamma(-\frac{\nu}2)}\Bigl(-2\log\left|\frac{b}{2R}\right|\\
&\qquad\quad +\psi\!\left(-\frac{\nu}2\right)+\psi\!\left(\frac{D-2+\nu}2\right)\Bigr)|b|^{2-D-\nu}\nn
\end{align}
with the digamma function $\psi(z) := \Gamma'(z)/\Gamma(z)$. This may be most easily
derived from the Fourier transform of $|q|^{\nu}$ via a derivative on the
exponent $\nu$.
At 4PM order the impulse separates into the
test-body contributions with linear mass dependence, $m_{1} m_{2}^{4}$ or   $m_{1}^{4} m_{2}$, and the
comparable-mass contributions, $m_{1}^{2} m_{2}^{3}$ or   $m_{1}^{3} m_{2}^{2}$.
In total we face 258 graphs contributing to the 4PM tidal effects --- see \Fig{fig:tidal-graphs1} for some examples.  
The divergences arise in the comparable mass sectors.
The post-adiabatic contributions involving the counter-term are depicted
in \Fig{fig:tidal-graphs2} and amount to a 1-loop integration. Due to the
$m^{3}$ factor in the counterterm vertices --- \Fig{fig:tidalvertices} --- they contribute
to $m_{1}^{2} m_{2}^{3}$ and $m_{1}^{3} m_{2}^{2}$ terms as well, thereby canceling the
$1/\eps$ poles.

\sec{Impulse} Tidal contributions to the 4PM impulse may be split into a conservative $\Delta p^{(4)\mu}_{i,\text{cons}}$ and dissipative contribution $\Delta p^{(4)\mu}_{i,\text{diss}}$ due to the presence
of radiative (R) or potential (P) bulk gravitons \cite{Jakobsen:2023ndj,Jakobsen:2023hig}.
At the 4PM level only two gravitons may go on-shell and can become radiative. 
The conservative sector is given by the (PP) region and also receives contributions
from the (RR) part --- these may be identified upon using Feynman propagators for the
bulk gravitons. Dissipative contributions, on the other hand, emerge from the
mixed (PR) contribution and the remainder of the (RR) part --- reflecting 
the number of radiative gravitons.

Let us begin with the post-adiabatic contributions to the 4PM impulse,  proportional
to $\kappa^{(i)}_{\dot E^{2}/\dot B ^{2}}(R)$ and involving one-loop integrals, 
of
Fig.~\ref{fig:tidal-graphs2}, that, for NS 1, take the form
\begin{align}
  &\Del p^{(4)\, \mu}_{\rm 1,tidal'}
  =
  \hat b^\mu \frac{1575 G^{4 }m_1^2 m_2^2}{512 |b|^8}
  \pi \gam v
  \sum_{X,i}
  f_{X^2}m_ic_{X^2}^{(i)}
  \kappa_{\dot X^2}^{(i)}(R)\,,
\end{align}
and which we label by a prime on the tidal subscript.
The index $i$ runs over the two particles and $X$ over $E$ and $B$ with functions $f_{X^2}(\gam)$ given by:
\begin{align}
  f_{E^2}
  &=
  21\gam^4-14\gam^2+9
  \,, \,\,\,
  f_{B^2}
  =
  7(3\gam^3-2\gam^2-1)\, ,
\end{align}
The adiabatic tidal contributions to the conservative impulse of NS $1$ takes the form
\begin{align}
\label{Deltapcons}
  &\Delta p_{1,\text{tidal,cons}}^{(4)\mu} = 
  \frac{G^{4}m_{1}^2m_{2}^2}{|b|^{8}}
  \sum_{l=1}^{3}
  \rho^{\mu}_{l}
  \bigg[
    \frac{m_{2}^{2}}{m_1} 
    C_{l}(\gamma) 
    +
    \frac{m_{2}^{2}}{m_1} 
    {\bar C}_{l}(\gamma) 
    \nn \\
    &\qquad+
    \sum_{\alpha =1 }^{19}
    F_{\alpha}(\gamma)
    \Big(
    m_2
    D_{ \alpha, l}(\gamma)
    +
    m_1
    \bar{D}_{\alpha, l}(\gamma)
    \Big)
    \bigg]
  \\
  &\qquad+
  \hat b^\mu\frac{1605G^{4}m_1^2 m_2^2}{128|b|^8}\pi\sqrt{\gam^2-1} \log\Big|\frac{2 b}{R}\Big|
  \sum_{i,X}
  f_{X^2} m_i c_{X^2}^{(i)}\,,\nn
\end{align}
where $\rho^{\mu}_{l}= \{ \hat b^{\mu} , v_{1}^{\mu}, v_{2}^{\mu} \}$.
The coefficient functions $C,\bar C, D$ and $\bar D$
are linear in the tidal Love numbers and rational functions of $\gamma$,
up to  integer powers of $\sqrt{\gamma^{2}-1}$
\begin{align}
\label{CDs}
C_{l}(\gamma) &=\sum_{i=1,2} c_{E^{2}}^{(i)} C^{(i)}_{E, l}(\gamma) + c_{B^{2}}^{(i)} C^{(i)}_{B,l}(\gamma) \, , \nn\\
D_{\alpha,l}(\gamma) &=\sum_{i=1,2} c_{E^{2}}^{(i)} D^{(i)}_{E, \alpha, l}(\gamma) + c_{B^{2}}^{(i)} D^{(i)}_{B,\alpha, l}(\gamma) \, .
\end{align}
Analogous relations hold for the barred quantities. We find 19 basis functions from
the three loop-integrals
at the 4PM order, which we choose to be even in
$v=\sqrt{1-\gamma^{-2}}$, of the form:
\allowdisplaybreaks
\begin{align}\label{eq:functionBases}
  &F_{1,\ldots, 5}
  \!=\!
  \bigg\{1,
  \dfrac{
    \log[x]}{\sqrt{\gamma^2-1}},
  \log\! \left[\frac{\gamma_{+}}{2}\right],
  \log^2[x],
  \dfrac{\log[x]\log\!\left[\frac{\gamma_+}{2}\right]}{\sqrt{\gamma^2-1}}
  \bigg\},
  \nn
  \\
  &F_{6,\ldots,9}=
  \bigg\{
  \log[\gamma],
  \log^2\! \left[\frac{\gamma_{+}}{2}\right],
  \text{Li}_{2}\left [\frac{\gamma_{-}}{\gamma_{+}}\right],
  \text{Li}_{2}\left [-\frac{\gamma_{-}}{\gamma_{+}}\right]
  \bigg\}\,,
  \nn
  \\
  &F_{10,\ldots,13}
  =
  \bigg\{
  \dfrac{\log[x]}{\sqrt{\gamma^2-1}}\log[\gamma],
  \frac{1}{\sqrt{\gamma^2-1}}\chi_2\left[\sqrt{\frac{\gamma_-}{\gamma_+}}\,\right],
  \nn
  \\
  &\qquad\qquad
  \text{Li}_2[-x^{2}] -4\text{Li}_2[-x] -\log[4] \log[x] 
  -\frac{\pi^{2}}{4}
  ,
  \nn
  \\
  &\qquad\qquad\qquad
  \dfrac{\text{Li}_2[-x]-\text{Li}_2\left[-\frac{1}{x}\right]+\log[4] \log[x]}
    {\sqrt{\gamma^2-1}}
    \bigg\}, \\
       &F_{14,15,16}
  =
  \bigg\{ 
     \eE^2\left [ \frac{\gamma_{-}}{\gamma_{+}} \right ],
   \eK^2\left [ \frac{\gamma_{-}}{\gamma_{+}} \right ],
   \eE\left [ \frac{\gamma_{-}}{\gamma_{+}} \right ]
   \eK\left [ \frac{\gamma_{-}}{\gamma_{+}} \right ]
    \bigg\}, \nn \\
       &F_{17,18,19}
    = \bigg\{
    \log \left[\frac{\gamma_{-}}{2} \right],
    \frac{ \log 
      \left [\frac{\gamma_{-}}{2} \right ]\log\left [x\right ]}{\sqrt{\gamma^{2}-1}},
    \nn\\&
  \qquad\qquad\qquad
  \log \left [\frac{\gamma_{-}}{2}\right ] \log\left [\frac{\gamma_{+}}{2} \right ]
  \bigg\}, \nn
\end{align}
where $\gamma_{\pm}=\gamma\pm 1$, $x=\gamma - \sqrt{\gamma^{2}-1}$
and $\chi_\nu[z]=\sfrac12({\rm Li}_\nu[z]-{\rm Li}_\nu[-z])$
is the Legendre chi function. Note the appearance of complete elliptic integrals of the first and  second kind in entries $F_{14,15,16}$. 
In the conservative impulse   the contributions to the $v_{i}^{\mu}$ directions, i.e.~$l=2,3$ in \eqref{Deltapcons},
only pickup the basis functions $F_{1}$ and $F_{2}$.

For the dissipative sector we find
\begin{align}
  \Delta p_{1,\text{tidal,diss}}^{(4)\mu} =& 
  \frac{G^4m_{1}^2m_{2}^2}{|b|^{8}}
  \sum_{l=1}^{3}  \rho^{\mu}_{l}\\ &
     \sum_{\alpha =1 }^{13}
    F_{\alpha}(\gamma)
    \Big(
    m_2
    E_{ \alpha, l}(\gamma)
    +
    m_1
    \bar{E}_{\alpha, l}(\gamma)
    \Big)
    \bigg]\,,\nn
\end{align}
which only uses the first 13 basis functions of \eqref{eq:functionBases} 
and does not have a probe-limit
$m_{1}m_{2}^{4}$ or $m_{1}^{4}m_{2}$ contribution.
As mentioned above, the (RP) and (RR) regions contribute here, which we denote
in the following as $\text{rad}^{1}$ and $\text{rad}^{2}$ respectively.
In analogy to \eqref{CDs} the coefficient functions read
\begin{align}
\label{Es}
E_{\alpha,l}(\gamma) &=\sum_{i=1,2} c_{E^{2}}^{(i)} E^{(i)}_{E, \alpha, l}(\gamma) + c_{B^{2}}^{(i)} E^{(i)}_{B,\alpha, l}(\gamma) \, ,
\end{align}
and similarly for the barred one. The explicit form of \eqref{CDs} and \eqref{Es}
are collected in the ancillary file included in the {\tt arXiv.org} submission of
this article. 

\sec{Scattering angle}  A relative, dissipative scattering angle $\theta$ may be defined as follows.
First, we define the relative momentum,
\begin{align}
  {p}^\mu = \frac{\nu}{\Gamma^2}
  \Big(
  \frac{\gam m_1+m_2}{m_1}
  p_1^\mu
  -
  \frac{\gam m_2+m_1}{m_2}
  p_2^\mu
  \Big)
  \ ,
\end{align}
with $\nu=m_{1}m_{2}/M^{2}$,  total mass $M=m_{1}+m_{2}$ and $\Gamma=E/M$,
such that in the initial center-of-mass (CoM) frame total
momentum $P^\mu=p_{1}^{\mu}+p^{\mu}_{2}$ we have $p_1^\mu = (E_1,\vct{p})$ and $p_2^\mu = (E_2,-\vct{p})$ with $p^\mu =(0,\vct{p})$ and $E=|P^{\mu}|$.
The relative scattering angle is now defined as the angle between the initial and final value of $p^\mu$ taken in the \textit{initial} CoM frame.
For planar scattering one finds the formula,
\begin{align}
  \tan(\theta)
  =
  -\frac{
    \hat b\cdot \Del p}{
    \pin - \hat p\cdot \Del p}
  \ ,
\end{align}
which for conservative scattering reduces to
\begin{align}
   \sin\Big(\frac{\theta_{\rm cons }}{2}\Big)
   =
   \frac{|\Delta p^\mu_{i,\rm cons}|}{2 p_\infty}
   \ .
\end{align}
The angle is PM-expanded, $\theta=\sum_{n=1}^\infty\theta^{(n)}$, and expanded in the tidal couplings,
\begin{align}
  \theta^{(n)}
  &=
  \theta^{(n)}_{\rm pp}
  +\theta_{\rm tidal'}^{(n)}
  +
  \sum_{X=E,B} 
  \Big(
  \theta_{X^2}^{(n,+)}
  c_{X^2}^{(+)}
  +
  \theta_{X^2}^{(n,-)}
  \delta
  c_{X^2}^{(-)}
  \Big)
  \,.
\end{align}
The first term describes the point-particle tidal-free part, the second term post-adiabatic tidal effects and the final terms adiabatic tidal corrections.
The relative mass difference is $\delta=(m_1-m_2)/M$ and we use symmetric finite-size couplings defined by:
\begin{align}
  c_{E^2}^{(\pm)}
  &=
  c_{E^2}^{(1)}\pm c_{E^2}^{(2)}
  \,, &
  c_{B^2}^{(\pm)}
  &=
  c_{B^2}^{(1)}\pm c_{B^2}^{(2)}
  \ .
\end{align}
The leading adiabatic tidal effects appear at second PM order and the leading post-adiabatic effects at fourth PM order (in our formal PM counting).

The (leading-order) 4PM post-adiabatic angle is derived from one-loop integrals, and reads
\begin{align}
  \theta^{(4)}_{\rm tidal'}
  =
  \Gamma \frac{1575\pi\nu}{512}
  \frac{(GM)^4}{|b|^8}
  \sum_{X}
  f_{X^2}\kappa^{(+)}_{\dot X^2}(R)
  \ ,
\end{align}
with
\begin{align}
  \kappa^{(+)}_{\dot X^2}(R)
  =
  \frac{m_1 c_{X^2}^{(1)} \kappa_{\dot X^2}^{(1)}(R)
    +
  m_2 c_{X^2}^{(2)} \kappa_{\dot X^2}^{(2)}(R)
  }{M}
  \ .
\end{align}
The 4PM tidal contributions take a similar form as the tidal-free angle:
\begin{align}
  \theta_{X^2}^{(4,\pm)}&=
  \Gam
  \frac{(GM)^4}{|b|^8}
  \Big[
  \theta_{\rm X^2\!,cons,\nu^0}^{(\pm)}
  +
  \nu \theta_{\rm X^2\!,cons,\nu^1}^{(\pm)}
  \\
  &\qquad\qquad\qquad+
  \frac{\nu}{\Gam^2}
  \Big(
  \theta_{\rm X^2\!,diss,\nu^1}^{(\pm)}
  +
  \nu\theta_{\rm X^2\!,diss,\nu^2}^{(\pm)}
  \Big)
  \Big]\,.
  \nn
\end{align}
The angle coefficients of this expansion depend only on $\gam$ and $\log | 2b/R|$,
and may be expanded on the function basis $F_\alpha(\gam)$ in terms of polynomials of $\gamma$ (up to $\sqrt{\gam^2-1}$).

The angle satisfies the same tail relation as pointed out in Ref.~\cite{Jakobsen:2023ndj}:
\begin{align}
  \theta_{\rm tidal,tail}^{(4)}
  =
  G E
  \frac{
    \pat E_{\rm tidal, rad}^{(3)}}{
    \pat L}
  \log\Big(
  \frac{\gam-1}{2}
  \Big)\,,
  \label{tail}
\end{align}
where we define $\theta_{\rm tidal, tail}^{(4)}$ as the part of $\theta_{\rm tidal}^{(4)}$ in the direction of the tail functions $F_{\alpha}$ with $\alpha=17,18,19$
that depend on $\log[\gamma_{-}/2]$, $E_{\rm rad}$ being the radiated energy.

\sec{Linear response}
The dissipative angle obeys the same linear response relation as derived in Ref.~\cite{Jakobsen:2023hig}
\begin{align}
  \theta_{\rm tidal,rad^1}^{(4)}
  =
  -
  \frac12
  \Big(
  &
  \frac{
    \pat\theta^{(1)}_{\text{pp}}}{
    \pat L}
  L_{\rm tidal,rad}^{(3)}
  +
  \frac{
    \pat\theta^{(1)}_{\text{pp}}}{
    \pat E}
  E_{\rm tidal,rad}^{(3)}
  \nn
  \\
  &+
  \frac{
    \pat\theta^{(2)}_{\rm tidal}}{
    \pat L}
  L_{\rm pp, rad}^{(2)}  
  \Big)
  \label{linearResponse}
\end{align}
where the ``pp'' subscript refers to point-particle (and so tidal-free) contributions.
We note that neither the 1PM angle nor the 2PM loss of angular momentum have a tidal contribution.
On the left-hand-side, the subscript ``rad$^1$'' refers to the part of the angle including a single radiative graviton.
This part may also be identified from its odd scaling under $v\to-v$.
Knowledge of the 4PM rad$^1$ angle completely determines the 3PM tidal loss of angular momentum,
and vice versa (assuming knowledge of other relevant 3PM observables).
The 3PM tidal loss of angular momentum was previously derived in Ref.~\cite{Heissenberg:2022tsn}
with which we fully agree.
The result for $L_{\rm rad}^{(3)}$ reads
\begin{align}
  L_{\rm tidal,rad}^{(3)}
  &=
  \frac{\pi G^3  M^4 \nu^2}{
    |b|^6\Gam^3}
  \sum_{\alpha=1}^3
  F_\alpha(\gam)
  \sum_{X=E,B}
  \\
  &\qquad
  \times\Big(
  H_{\alpha,X^2}^{(+)}(\gam)
  c_{X^2}^{(+)}
  +
  H_{\alpha,X^2}^{(-)}(\gam)
  \delta
  c_{X^2}^{(-)}
  \Big)\nn
  \ ,
\end{align}
where the functions $H_{i,X^2}^{\pm}(\gam)$ are polynomial (up to $\sqrt{\gam^2-1}$) in $\gam$ and is a linear function in $\nu$.
This result, together with point-particle result, is provided in the ancillary file 
on {\tt arXiv.org}.

\sec{Checks}
The results in this paper build on the 3PM results of Ref.~\cite{Jakobsen:2022psy}, and naturally agree with them.
In addition, we have checked that the post-Newtonian limit $v\to0$ of the conservative scattering angle reproduces the 2PN and 3PN scattering angles reported in 
Refs.~\cite{Mandal:2023lgy,Mandal:2023hqa}.\footnote{Modulo a typo in eq.~(6.14) of that paper (v1): the mass ratios should be dropped. We thank the authors for communication.}
They also obey the non-trivial checks Eqs.~\eqref{tail} and~\eqref{linearResponse} (reproducing the 3PM loss of angular momentum of Ref.~\cite{Heissenberg:2022tsn}) and in general we have checked that the impulse obeys the constraints $(p_i+\Del p_i)^2=p_i^2$, which
provides a further internal consistency check.

\sec{Conclusions} In this Letter we have applied the worldline quantum field theory
formalism to tidal effects at 4PM order, demonstrating the power of our technology. It is worth
stressing that the work flow was identical (if not simpler) than in the case of
spin~\cite{Jakobsen:2023ndj,Jakobsen:2023hig}. We established tidal effects in the impulse in the conservative and dissipative 
sectors at 4PM  and derived the conservative and dissipative scattering angle. A new feature
appearing at this NNLO order is the need to include post-adiabatic couplings in order to cancel a 
divergence in this classical field theory computation which results in a renormalization
group flow of the post-adiabatic Love numbers.
We also confirmed the tidal contributions to
the radiated angular momentum previously derived by Heissenberg via very different 
methods \cite{Heissenberg:2022tsn}. Our findings will be potentially useful for
improving future waveform models to include tidal effects \cite{Hinderer:2016eia}. They represent yet another mosaic stone in our steadily
improving picture of highest-precision gravitational wave physics.

\medskip

\sec{Acknowledgments}
We thank Tanja Hinderer, Rafael Porto, Muddu Saketh and especially Carlo Heissenberg, Raj Patil and Jan Steinhoff 
for discussions.
This work was funded by the Deutsche Forschungsgemeinschaft
(DFG, German Research Foundation)
Projektnummer 417533893/GRK2575 ``Rethinking Quantum Field Theory''
and  by the European Union through the 
European Research Council under grant ERC-AdG-101097219 (GraWFTy). 
Views and opinions expressed are however those of the authors only and do not necessarily reflect those of the European Union or European Research Council Executive Agency. Neither the European Union nor the granting authority can be held responsible for them.

\bibliographystyle{JHEP}
\bibliography{tidal-4PM}

\end{document}